\documentstyle[12pt]{article}
\begin{document}
\title{Hadronic production of heavy mesons in perturbative QCD}
\author{A.V.~Berezhnoy, V.V.~Kiselev, A.K.~Likhoded}
\date{~}
\maketitle
\begin{center}
{\small Institute for High Energy Physics,\\
Protvino, Moscow Region, 142284, Russia.\\
E-mail: likhoded@mx.ihep.su, ~~Fax: (095)-230-23-37}
\end{center}
\vspace*{3mm}

\begin{abstract}
In the framework of perturbative QCD in the fourth order over the $\alpha_s$
coupling constant for the production of two pairs of quarks and in the
model of weak binding of the quarks into meson, the analysis of the hadronic
production of mesons, containing $b$-quark, is performed, and a minimal
transverse momentum is determined, so that at the momenta greater than the
found one, the meson differential spectrum can be reliably described 
in the model of factorization of the hard $b$-quark production and 
the subsequent fragmentation into the mesons. At low transverse momenta
of the meson, nonfragmentational contributions are essential. They are
determined by the complete set of QCD diagrams in the given order over
$\alpha_s$, so that the latters result in the effect of destructive
interference with the diagrams of fragmentation at
$z=2E/\sqrt{s}$ close to $1$.
\end{abstract}

\section{Introduction}

In the study of the heavy quark production mechanisms, the consideration within
the perturbative theory is basic. This is related with the fact that
the heavy quark mass $m_Q$ determines the scale being much greater than
the confinement scale in QCD, $m_Q \gg \Lambda_{QCD}$. Then the cross-section
of the heavy quark production is factorized as the partonic cross-section
of the hard production of $Q\bar Q$ quarks and the distribution functions of 
quarks and gluons, in the interaction of which the $Q\bar Q$ pair is produced.

Next step is connected with the heavy quark transform into hadrons containing
these quarks. Ordinary in this consideration, one uses the model of the heavy
quark fragmentation into the quarkonium $Q\to (Q\bar q)+X$ with the
fragmentation functions measured in the $(Q\bar q)$-hadron production in 
$¥^+e^-$-annihilation, for instance. Such approach is the generalization of
the factorization theorem on the case of hadronic production \cite{soper}.
Apparently, the scheme mentioned above does not satisfactory work as one can 
see, say, in the $B$-meson production at the FNAL collider energy \cite{cdf}.
Indeed, the result of theoretical predictions for the $b$-quark production 
in the $O(\alpha_s^3)$-calculations is in the systematic disagreement with
the experiment. On the other hand, the production of mixed flavor heavy
quarkonium, $B_c$, considered in refs. \cite{yaf,chen,ruckl}, where the
hadronization is already included as the composite element of the model,
points out the deviation from the naive intuitive expectation of the
fragmentation regime dominance.

The detailed consideration of the $B_c$-meson production processes shows, 
for instance, that in a broad kinematical region these processes can not
be reduced to the simple consequence: 1) the $b$-quark production, 2) the
$b$-quark fragmentation into the $B_c$-mesons with the fragmentation function
$D_{b\to B_c}(z)$ known from the calculations in $e^+e^-$-annihilation.
So, for the $B_c$ production in $\gamma \gamma $-collisions considered in
\cite{plet}, the $c$-quark fragmentation into $B_c$ is greater than the
$b\to B_c$ fragmentation. Remember, that in $e^+e^-$-annihilation the
$c\to B_c$ fragmentation is suppressed by two orders of magnitude with respect
to the $b\to B_c$ fragmentation \cite{zphys}.

In the hadronic production, where there is no such enforcement of the $c$-quark
contribution due to the $(Q_c/Q_b)^4=16$ factor, the $b$-quark fragmentation
into $B_c$ is still not dominant. The naive expectations of the fragmentation
dominance are just realized at $p_T\gg M_{B_c}$ ($p_T > 30$ GeV for $B_c$ and 
$p_T > 40$ GeV for $B_c^*$). By the way, the contribution of diagrams
different from those of the fragmentation type on the topology
(the recombination contribution in the rest of the paper) is basic.

In this paper we investigate in what measure these additional contributions 
can be essential in the $B$-meson production. We evaluate the transverse
momentum, so that at momenta greater than the determined one, the heavy 
meson production can be reliably described by the fragmentation model.
We study these problems in the framework of model for the production of
weakly bound quarkonium with the following standard assumptions.
1) The calculation of the production cross-section for two pairs of quarks
is performed in the leading Born approximation (diagrams of the fourth 
order over $\alpha_s$). 2)  We neglect the binding energy and relative 
quark motion inside the meson. The latter means that the quarks inside the
$(Q\bar q)$-meson move with the parallel momenta
$p_{\bar q}=p_Q\frac{m_q}{m_Q}$. Since the derivation of analytic expressions
for the considered processes looks 
to be problematic, we use a numerical way for
the evaluation of amplitudes corresponding to 36 Feynman diagrams in the
hadronic production and to 20 diagrams in the photonic production of 
quarkonium.

\section{Factorization of inclusive spectrum of heavy meson}

The basic subject of our discussion below will be the factorization
of the inclusive cross-section for the hadronic production of heavy 
meson in the region of high transverse momenta. In present Section
we consider the subprocess of the gluon-gluon production of the meson.
The physical basis for the construction of the factorization model is 
the appearance of two energetic scales determining the hard production 
of quarks at high transverse momenta and a soft nonperturbative forming of
the bound state. The case, when the conditions 
$\sqrt{s}\gg m_Q\gg m_q\gg \Lambda_{\rm QCD}$ are valid, is of a special 
interest. At the given ordering of scales, the quark production allows the
description in the framework of perturbative QCD, and their hadronization into
the meson is described by the nonrelativistic model of weakly bound state:
the energy of the quark binding is low, $\epsilon_{Q\bar q}\ll m_q,\; m_Q$,
and the perturbative production of quarks takes place in the region, where
the quarks composing the meson move with the same velocity $v_\mu$
equal to the meson velocity. Then the quark momenta are equal to
$p_Q^\mu= m_Q\cdot v^\mu\;,\;\; p_q^\mu= m_q\cdot v^\mu$. The
$m_Q\gg m_q$ condition provides the appearance of two scales in the
production process of quarks with different flavors, so that we can talk on the
$Q$-quark fragmentation into the $(Q\bar q)$-meson. 

The technique of such calculations was given in refs. \cite{yaf,plet,zphys}, 
in details. 

As in the model of fragmentation, in the $O(\alpha_s^4)$-order over the
QCD coupling constant the cross-section of the subprocess
$gg \to (Q\bar q) + \bar Q + q$ is proportional to the
$\alpha_s^4 |\Psi(0)|^2$ value, which determines an overall normalization of 
the cross-section and it includes all numeric effects of large distances
($\Lambda_{\rm QCD}$ and the meson size $<r_{Q\bar q}>$). In concrete 
calculations we have supposed 
$m_Q/m_q\simeq 17$, $\sqrt{s}/m_Q\simeq 20$, $m_q/\Lambda_{\rm QCD}\simeq 10$.

By the general theorem on factorization \cite{soper} it is clear that
at high transverse momenta the fragmentation of the heavier quark
$Q\to (Q\bar q) + q$ must dominate. It is described by the factorized 
formula
\begin{equation}
\frac{d\sigma}{dp_T} = \int \frac{d\hat \sigma(\mu; gg\to Q\bar Q)}
{dk_T}_{|_{k_T=p_T/x}}\cdot D^{Q\to (Q\bar q)}(x;\mu)\; \frac{dx}{x}\;,
\label{one}
\end{equation}
where $\mu$ is the factorization scale, $d\hat\sigma/dk_T$ is the cross-section
for the gluon-gluon production of quarks $Q+\bar Q$, $D$ is the fragmentation
function. In what follows, we consider the production of the $S$-wave
states in the $(Q\bar q)$ system: the vector  $1^-$ state and 
pseudoscalar $0^-$ one. For these states the scaling analytic expressions
of the fragmentation functions were derived at $M^2/s\ll 1$ in the leading 
order of perturbative QCD \cite{zphys,chin1,braat}.

The distributions over the meson transverse momentum in the gluon-gluon 
production are shown in Fig. \ref{fig1}, in which one can see that at 
$p_T> p_T^{\rm min} \gg M_{(Q\bar q)}$ the exact perturbative
calculations of the $O(\alpha_s^4)$-order prove the validity of the
factorization theorem, and they fix the scale, from which the production
mechanism  reaches the regime of fragmentation. Note, that numerically 
at the chosen ratios of the quark masses and the total energy of the partonic
subprocess we have $p_T^{\rm min} \sim (5 \div 6) \cdot m_Q$, and the latter
value is greater for the vector state. The analysis shows also that the low
boundary for the applicability of the factorization approach shifts to the
region of lower momenta at the decrease of the $m_q/m_Q$ ratio.

Compare now the result of the exact perturbative $O(\alpha_s^4)$-calculation 
for the differential cross-section $d\sigma/dz$, $z=2 E/\sqrt{\hat s}$,
with the result of the factorization model analogous to (\ref{one}). 
Remember, that in $e^+e^-$-annihilation the distribution under consideration 
coincides with the fragmentation function. One can see in Fig. \ref{fig2},
that, in contrast to the fragmentation model, there is a sizable additional
contribution of the recombination type in the region of low $z$ 
(i.e. at low energies and, hence, low $p_T$). We have found that the
recombination contribution is generally given by the diagrams of the 
evolution type for the gluon splitting into the $q\bar q$ or $gg$ pairs. 
However, the advantage of the given perturbative approach at 
$m_q\gg \Lambda_{\rm QCD}$ is the exact, process dependent account for
the finite values of masses, virtualities and transverse momenta, in contrast
to the universal approximate method of Altarelli--Parisi.

We have in details considered the contributions of
each diagram in the region of $z\to 1$. In the covariant Feynman gauge the
diagrams of the gluon-gluon production of $Q+\bar Q$ with the subsequent
$Q\to (Q\bar q)$ fragmentation dominate as well as the diagrams,
when the $q\bar q$ pair is produced in the region
of the initial gluon splitting, are large. 
However, as one can see in Fig. \ref{fig2}, 
the contribution of the latters diagrams leads to the destructive interference 
with the fragmentation amplitude, and this results in the "reduction"
of the production cross-section in the region of $z$ close to 1. So,
$\Delta\sigma/\sigma$ is about 50\%. In the axial gauge with the vector
$n^\mu=p^\mu_{\bar Q}$ this effect of the interference has still a more bright 
appearance, since the diagrams like the splitting of gluons 
dominate by several orders of magnitude over the fragmentation, but the
destructive interference results in the cancellation of such extremely
large contributions. By the way, we have to note that discussions and 
speculations given in ref. \cite{chen} are certainly misleading, since the
authors draw conclusions by the consideration of the singularities in the
virtual particle propagators, which could determine a dominating contribution.
As we have just seen, some separately large contributions can destructively
interfere to cancel each to other. This interference is caused by the 
nonabelian nature of QCD, 
i.e. by the presence of the gluon selfaction vertices.
The comparison with the abelian case will be in details given in next Section.

\section{Gluon production of $B_c$}

The model for the production of the mixed flavor heavy quarkonium, as it
stands in previous Section, is the most realistic in the description of the
differential spectra for the $B_c$-mesons \cite{ger}, for which, however,
the $m_c/m_b$ ratio can not be yet considered as the small parameter.

As it was shown  in \cite{yaf,chen,ruckl} and one can see in Fig. \ref{fig3},
the fragmentation regime for the $B_c$-mesons is strongly delayed into 
the region of high transverse momenta, $p_T^{\rm min}\sim 35\div 40$ GeV.
The last fact is related with the comparatively large value of the $c$-quark
mass. Thus, one can again talk on the confirmation of the factorization theorem
for the hard gluonic production of $b$-quarks at high transverse momenta
$p_T\gg M_{B_c}$ and the less hard fragmentation of $b$-quark into 
$B_c^{(*)}$-meson ($m_c\ll p_T$). However, the large numeric value of 
$p_T^{\rm min}$ points out the fact that the basic amount of events of the 
hadronic $B_c^{(*)}$-production do not certainly allows the description in
the framework of the fragmentation model. This conclusion looks more evident, 
if one considers the $B_c$-meson spectrum over the energy 
(see Fig. \ref{fig5}).

The basic part of events for the gluon-gluon production of
$B_c$ is accumulated in the region of low $z$ close to 0, where the 
recombination being essentially greater than the fragmentation, dominates.
This is the significant difference between the $B_c$ production and the
$(Q\bar q)$-meson one, considered in previous Section, where both the 
fragmentation and recombination have given comparable contributions. As it
stands for $m_Q\gg m_q$, one can draw the conclusion on the  essential 
destructive interference in the region of $z$ close to 1, for the pseudoscalar
state.

To stress the role of the interference diagrams related to the nonabelian 
selfaction of gluons, we have considered the process with abelian currents. 
To avoid effects of an enforcement for the fragmentation contributions by 
different quarks due to the difference between their charges \cite{plet},
we have supposed the abelian charges of produced quarks to be equal each 
to other. As one can see in Fig. \ref{fig5}, where the spectrum over the energy
of the pseudoscalar state is shown, in the abelian case the effect of the 
destructive interference due to the additional contribution by the selfaction
of gauge quanta, is absent. So, the agreement between the factorized
model of fragmentation and the exact perturbative calculation is quite good at
$z$ close to 1.

The direct verification of the given mechanism for the $B_c$-meson production
could be the comparison of the $B_c$-meson spectra in two semispheres in 
the region of the gluon fragmentation and in the photon one, in the 
photonic production of $B_c$ at nucleons.

\section{Hadronic production of $B$-mesons}

The model of the gluon-gluon production of quarkonium, as it is described 
in Section 2, can be, after some notes, expanded to the hadronic production 
of mesons with a heavy quark $Q$ ($m_Q\gg \Lambda_{\rm QCD}$) and a light 
antiquark $\bar q$ ($m_q\sim \Lambda_{\rm QCD}$). Then one can talk on 
the model of meson composed of the constituent quarks, whose motions 
inside the meson are neglected, and, hence, one can apply the nonrelativistic 
model of quarkonium with weakly bound constituents. As for the 
nonperturbative parameters of model, $m_q$ and $|\Psi(0)|$, 
they can be phenomenologically fixed over the spectrum form and total
cross-section of the $B$-meson production in $e^+e^-$-annihilation at the
$Z$-boson peak, where the fragmentation regime dominates certainly.
Considering the $\sigma^{-1} d\sigma/dz$ spectrum over $z=2E/\sqrt{s}$,
one has to take into account the perturbative evolution breaking
the scaling for the function of the $b$-quark fragmentation into the $B$-meson,
$D^{b\to B}(\mu=\sqrt{s};z)$. After that, the $m_q=0.3$ GeV mass value agrees 
with the observed form of the fragmentation function, still the accuracy
of such estimate is low because of large experimental errors in parameters.
So, in the one-loop approximation for the perturbative evolution the average 
fraction of energy taken by the $B$-meson, equals
\begin{equation}
\langle z(\sqrt{s})\rangle = \biggl(\frac{\ln (m_b/\Lambda_{\rm QCD})}
{\ln (\sqrt{s}/\Lambda_{\rm QCD})}\biggr)^n\; \langle z(m_b)\rangle\;,
\label{two}
\end{equation}
where $n= 32/(9b)$, $b=11-2n_f/3$, $n_f=5$,
and $\langle z(m_b)\rangle$ is the scaling value of the momentum fraction 
determined over the $b\to B^{(*)}$ fragmentation functions
derived in \cite{zphys,chin1,braat}. From (\ref{two}), the experimental value
\cite{pdg}
$$
\langle z_B(m_Z)\rangle = 0.708 \pm 0.003 \pm 0.015\;,
$$
and $\Lambda_{\rm QCD}=85\pm 20$ MeV corresponding to
$\alpha_s(m_Z)=0.117\pm 0.005$ \cite{pdg} in the one-loop approximation, one
finds $m_q=0.2\div 0.3$ GeV. In the following calculations we suppose
$m_q=0.3$ GeV. As for the normalization of the total cross-section of the
$B$-meson production at the $Z$-boson peak, it is determined by the
$\alpha_s^2 |\Psi(0)|^2$ value, where $\Psi(0)$ is expressed through the
leptonic constant of the $B$-meson in the limit of static $b$-quark,
$\tilde f_B^{\rm stat}$,
$$
|\Psi(0)| = \tilde f_B^{\rm stat}\; \sqrt{\frac{M_B}{12}}\;.
$$
In various papers \cite{f} one gives estimates corresponding to 
$\tilde f_B^{\rm stat}= 200\div 320$ MeV. Supposing the relative yields of 
$B^0$, $B^+$ and $B_s^0$-mesons to be equal to $0.4$, $0.4$ and $0.2$,
we get $\alpha_s=0.8\pm 0.2$. Of course, this phenomenological value of 
QCD coupling constant is large for one can talk on the reliability
of perturbative calculations of the $B$-meson spectrum in the model of 
Section 2. Nevertheless, one can believe that, being phenomenological,
such model is able to describe the basic differential characteristics 
in the hadronic production of $B$-mesons, as it is in $e^+e^-$-annihilation.

In Section 2 we have chosen the quark mass ratios such that they correspond to
$m_Q=m_b=5.0$ GeV, $m_q=0.3$ GeV and $\sqrt{s}=100$ GeV.
The only difference between the real case of the $B$-meson production and 
the consideration given in Section 2, is the change of the overall normalization
of the total cross-section. The ratios of spectra over the $B^*$ and $B$-meson
transverse momenta are presented in Fig. \ref{fig6}, where
they are calculated in the model of fragmentation and in the
$O(\alpha_s^4)$-order of QCD in $p\bar p$-collisions at $\sqrt{s} =1.8$ TeV
with the use of gluon distributions given in \cite{CTEQ}. We see that the
cross-section has the significant additional contribution of recombination in
the region of $p_T< 15$ GeV, that qualitatively agrees with the experimental 
data from CDF \cite{cdf} (see Fig. \ref{fig7}).

One has to note that the gluonic luminosity of the $gg\to B+\bar b+ q$
process enforces the contribution of  the region near the threshold, 
$\sqrt{s_{th}} = 2(m_Q+m_q)$, where, strictly speaking, the fragmentation 
approach for the $b$-quarks at low momenta has uncertainties of the
principal character, since the physical phase space of the subprocess is 
suppressed in comparison with the two-particle phase space of free
$b$-quark pair.  The enforcement of the near-threshold region due to the
luminosities and phase space leads to the fact that the fragmentation model
evidently gives overestimating values for the differential cross-sections 
at low $p_T$. 

The uncertainty with the phase space decreases rapidly, if one
limits the total energy of the gluon subprocess from the down. So,
at $\sqrt{\hat s}> 60$ GeV one can clear see in Fig. \ref{fig6}, that
the fragmentation picture of the $B$-meson production does not correspond
to the exact perturbative $O(\alpha_s^4)$-calculations of the meson spectrum
in the model of hadronization for the weakly bound constituent quarks, at
$p_T< 25$ GeV. It is clear that at high energies, the multiple production of
light quark pairs can take place in addition to the single one.
The large value of the phenomenological constant $\alpha_s\approx 0.8$
points out the latter possibility. In the fourth order over $\alpha_s$
this means that the invariant mass of the $(b\bar q)$ pair can be large
in accordance with the $(q\bar q)$ pair mass can be close to the threshold.
Such contribution could make softer the $b$-quark spectrum, i.e. it could 
result in the additional contribution into the region of low $p_T < 15$ GeV, 
where the fragmentation regime does not still work.

Thus, we have shown that the perturbative $O(\alpha_s^4)$-calculations of QCD,
according to the reasonable model of the $b$-quark hadronization, confirm
the applicability of the factorization theorem at $p_T^B>15$ GeV.
They point out the appearance of additional, nonfragmentational 
contributions at low transverse momenta of the meson, $p_T^B<15$ GeV, 
so that these corrections appropriately reflect the observed experimental 
disagreement with the fragmentation model.

\section{Conclusion}

In the framework of perturbative QCD up to $O(\alpha_s^4)$-terms and in 
the simplest model of quark hadronization, we have analyzed the differential 
cross-sections of the heavy mesons over the transverse momenta and the 
energies. This analysis shows the following.

1) As one could expect by the theorem of factorization for hard processes,
in the region of $p_T>p_T^{\rm min}\gg M_{(Q\bar q)}$ the fragmentation
contribution dominates. At $p_T<p_T^{\rm min}$ the quark recombination into 
the heavy meson is basic, so that its relative value depends on the
$m_Q/m_q$ ratio. One has $p_T^{\rm min} \approx 40$ GeV for the $B_c$-mesons,
and the recombination contribution in the total cross-section is much greater 
than the fragmentation. One has $p_T^{\rm min}\approx 15$ GeV for the $B$-meson
production in hadron collisions, and the relative contributions by 
both the recombination and fragmentation are comparable with each other.

2) The exact perturbative $O(\alpha_s^4)$-calculations point out
the nonfragmentational contributions, which were not yet taken into 
account elsewhere previously, and they dominate in the region of low
$z=2E/\sqrt{s}$. These contributions increase by their relative magnitude 
in the total cross-section versus the decrease of the $m_Q/m_q$ ratio.

3) In the region of low transverse momenta and at $z$ close to 1, there is
the destructive interference of the fragmentation type diagrams with the 
diagrams of the initial gluon splitting into the $q\bar q$ pair,
so that the presence of the interference effect is determined by the 
nonabelian selfaction of gluons in QCD, and the effect is absent for the 
abelian currents.

This work is in part supported by the grant of Russian Foundation of
Fundamental Researches, $N^{\underline{0}}$~~ 96-02-18216.

\newpage
\centerline{\large \bf Figure Captions}
\begin{enumerate}
\item
The distributions over the transverse momenta of pseudoscalar and vector
$(Q\bar q)$-states, described in the $O(\alpha_s^4)$-order of QCD
(the dashed and solid line  histograms, correspondingly) and in the
fragmentation model (dashed and solid curves) in gluon-gluon collisions.
The cross-section is given in arbitrary units, the transverse momentum is 
in masses of $m_Q/5.$
\label{fig1}
\item
$d\hat \sigma/dz$ for the $(Q\bar q)$-system (notations as in Fig. 1).
\label{fig2}
\item
$d\hat \sigma/dp_T$ for the $B_c^{(*)}$-mesons at $\sqrt{\hat s}=200$ GeV
(notations as in Fig. 1).
\label{fig3}
\item
$d\hat \sigma/dz$ for the $B_c$-mesons at $\sqrt{\hat s}=100$ GeV.
The dashed histogram presents the $gg\to B_c +\bar b+ c$ process, the dotted
one is the abelian case. The curve shows the result of the fragmentation model.
The cross-sections are normalized over the cross-section of the $b\bar b$ pair
production.
\label{fig5}
\item
The ratio of spectra for the summed contributions of $B^*$ and $B$-mesons in
$p\bar p$-collisions at $\sqrt{s}=1.8$ TeV, as they are calculated in the
$O(\alpha_s^4)$-model of the weakly bound quarkonium and in the fragmentation 
model. $\sigma_{B_c^{(*)}}/\sigma^{\rm frag}_{B_c^{(*)}}(p_T)$ is given as the
solid histogram, the dashed histogram is the same but with the cut off
$\sqrt{\hat s}> 60$ GeV.
\label{fig6}
\item
The CDF data on the differential cross-section of the $B$-meson production at 
the FNAL Tevatron, in comparison with the $O(\alpha_s^3)$-prediction of
perturbative QCD for the $b$-quark production, 
$\sigma_{\rm exp}/\sigma_{\rm QCD}$.
\label{fig7}

\end{enumerate}
\newpage
\thispagestyle{empty}
\setlength{\unitlength}{1mm}
~~~~~~~~~~~~~~

\begin{figure}[t]
\begin{center}
\begin{picture}(100,110)
\put(10,10){\framebox(80,80){}}
\put(25,10){\line(0,1){2}}
\put(40,10){\line(0,1){2}}
\put(55,10){\line(0,1){2}}
\put(70,10){\line(0,1){2}}

\put(9,4){$6$}
\put(24,4){$10$}
\put(39,4){$14$}
\put(54,4){$18$}
\put(69,4){$22$}
\put(78,4){$p_T$, GeV}

\put(10,30){\line(1,0){2}}
\put(10,50){\line(1,0){2}}
\put(10,70){\line(1,0){2}}

\put(4,29){$1$}
\put(4,49){$2$}
\put(4,69){$3$}
\put(10,30){\line(1,0){80}}
\put(10,95){$\sigma_{\rm exp}/\sigma_{\rm QCD}$}

\put(14,72){\circle*{2}}
\put(25,42){\circle*{2}}
\put(39,50){\circle*{2}}
\put(63,38){\circle*{2}}

\put(14,58){\line(0,1){28}}
\put(25,34){\line(0,1){16}}
\put(39,42){\line(0,1){16}}
\put(63,30){\line(0,1){16}}

\end{picture}
\end{center}
\end{figure}
\vspace*{5mm}
\centerline{\large\bf Fig. \ref{fig7}}

\begin{thebibliography}{**}
\bibitem{soper}
J.C.~Collins, D.E.~Soper, Ann. Rev. Nucl. Part. Sci. 37, 383 (1987).
\bibitem{cdf}
F.~Abe et al., CDF Coll., FERMILAB-PUB-95/048-E (1995).
\bibitem{yaf}
A.V.~Berezhnoy, A.K.~Likhoded, M.V.~Shevlyagin, Yad. Fiz. 58, 730 (1995);\\
A.V.~Berezhnoy, A.K.~Likhoded, O.P.~Yuschenko, Preprint IHEP 95-59, Protvino
(1995) [hep-ph/9504302].
\bibitem{chen}
C.-H.~Chang et al., Phys. Lett. B364, 78 (1995).
\bibitem{ruckl}
K.~Kolodziej, A.~Leike, R.~R\" uckl, Phys. Lett. B355, 337 (1995).
\bibitem{plet}
A.V.~Berezhnoy, A.K.~Likhoded, M.V.Shevlyagin, Phys. Lett. B342, 351 (1995);\\
K.~Kolodziej, A.~Leike, R.~R\" uckl, Phys. Lett. B348, 219 (1995).
\bibitem{zphys}
V.V.~Kiselev, A.K.~Likhoded, M.V.~Shevlyagin, Z. Phys. C63, 77 (1994).
\bibitem{chin1}
C.-H.~Chang, Y.-Q.~Chen, Phys. Rev. D46, 3845 (1992), D50, 6013(E) (1994).
\bibitem{braat}
E.~Braaten, K.~Cheung, T.C.~Yuan, Phys. Rev. D48, 4230 (1993).
\bibitem{ger}
S.S.~Gershtein et al., Phys. Rev. D51, 3613 (1995);\\
S.S.~Gershtein et al., Uspekhi Fiz. Nauk 165, 3 (1995);\\
E.~Eichten, C.~Quigg, Phys. Rev. D49, 5845 (1994).
\bibitem{pdg}
L.~Montanet et al., PDG, Phys. Rev. D50, 1173 (1994);\\
M.~Feindt, Preprint CERN-PPE/95-76 (1995).
\bibitem{f}
M.~Neubert, Phys. Rep. 245, 259 (1994);\\
V.V.~Kiselev, Phys. Lett. B362, 173 (1995).
\bibitem{CTEQ}
J.~Botts et al., CTEQ Coll., Preprint ISU-NP-92-17, MSUHEP-92-27 (1992)

\end{thebibliography}
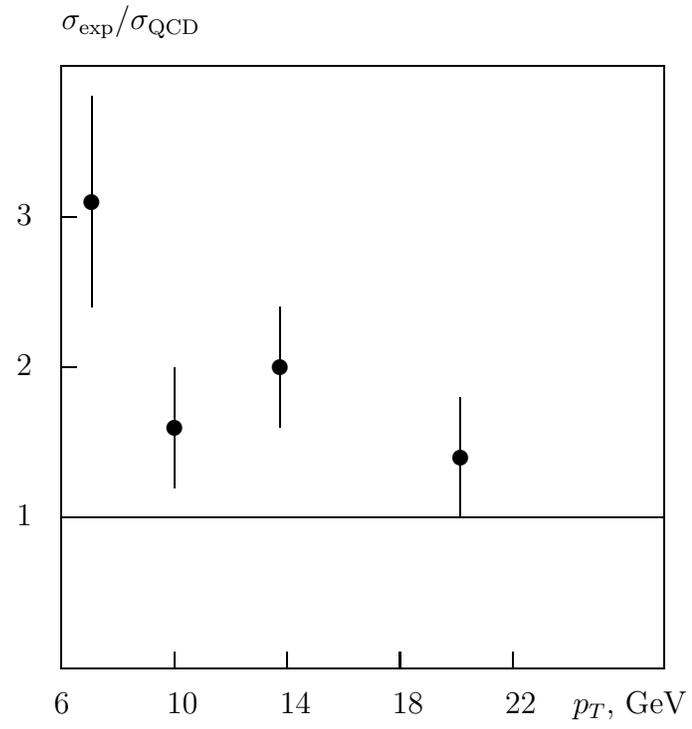
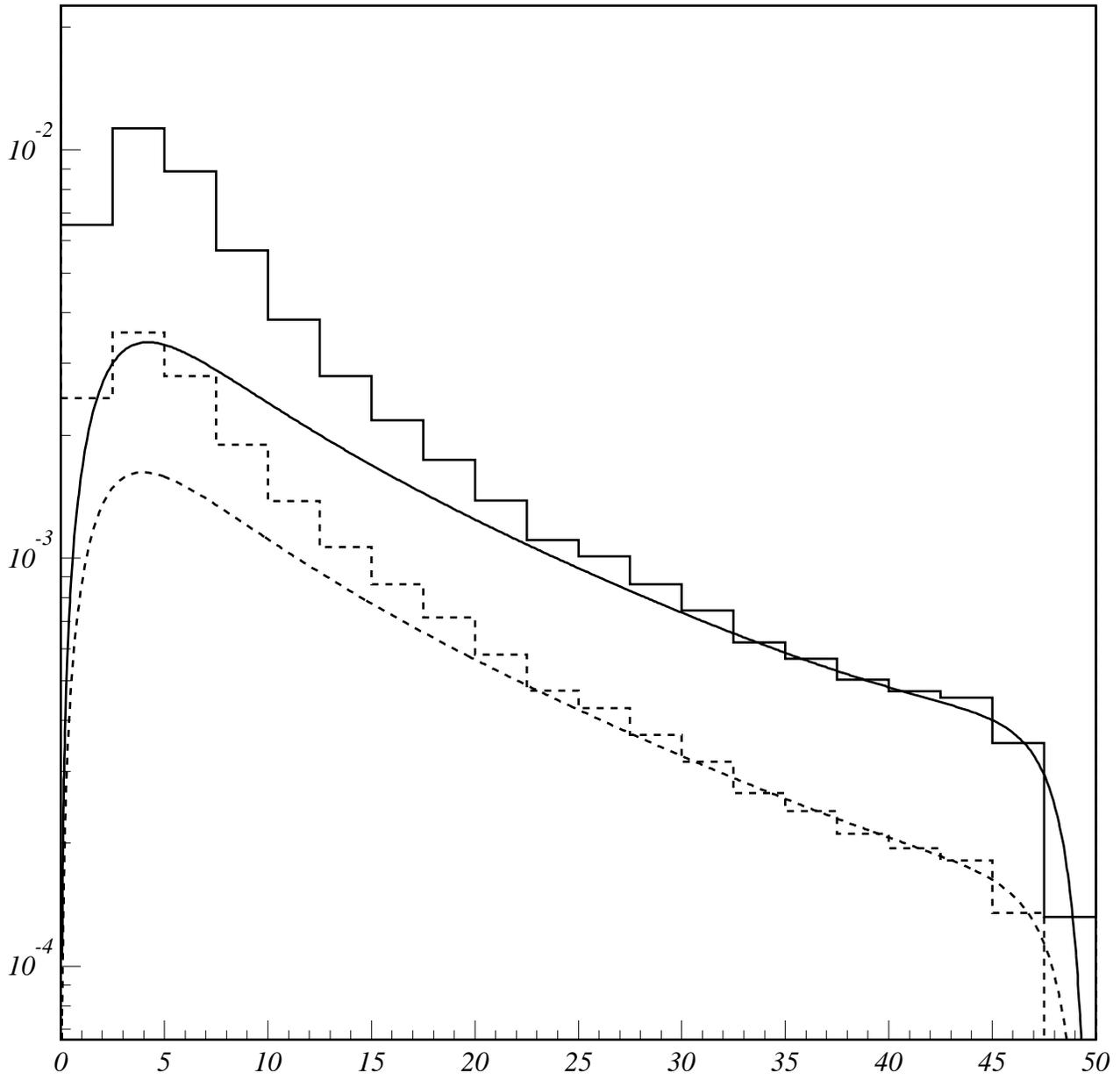
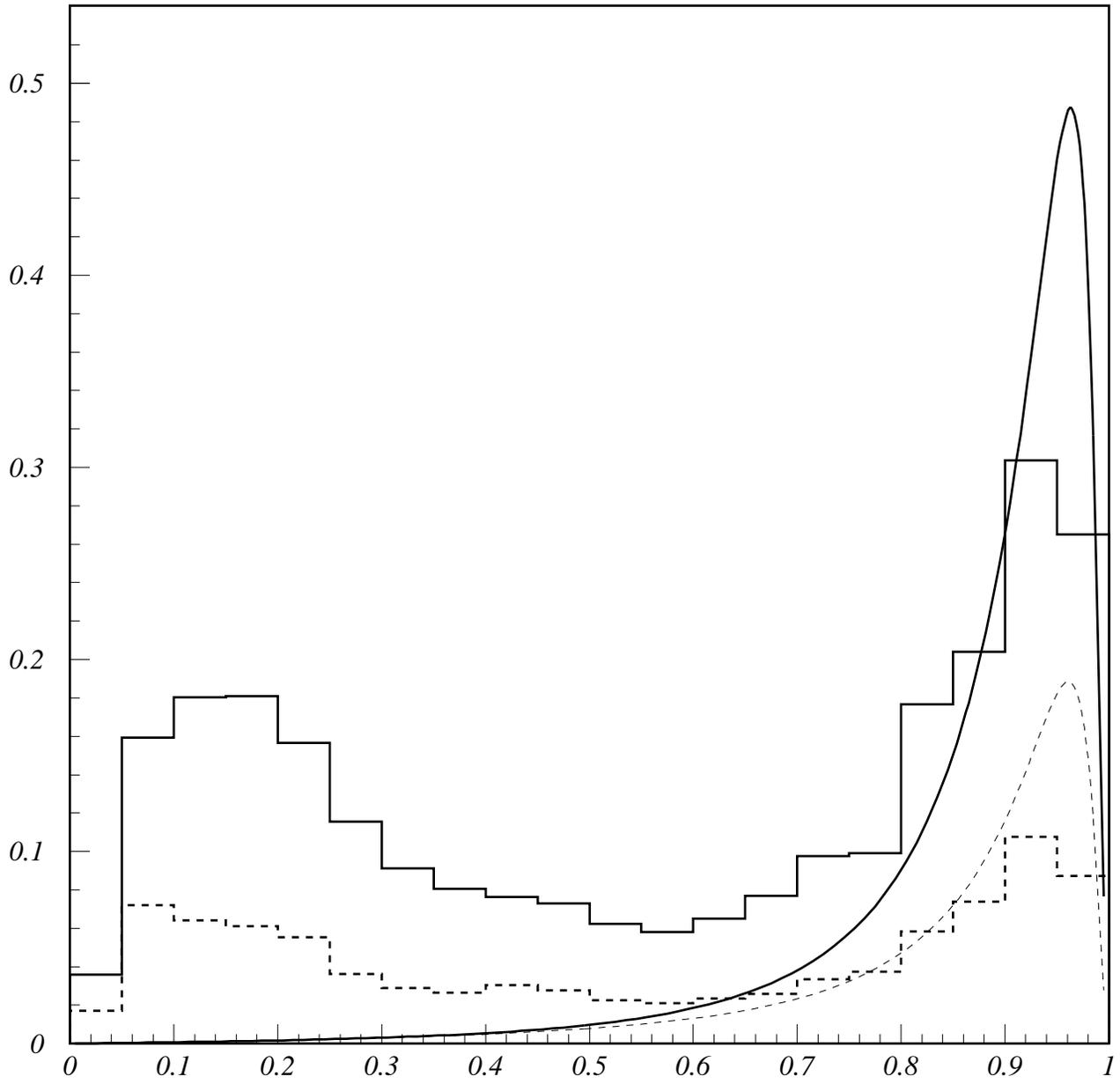
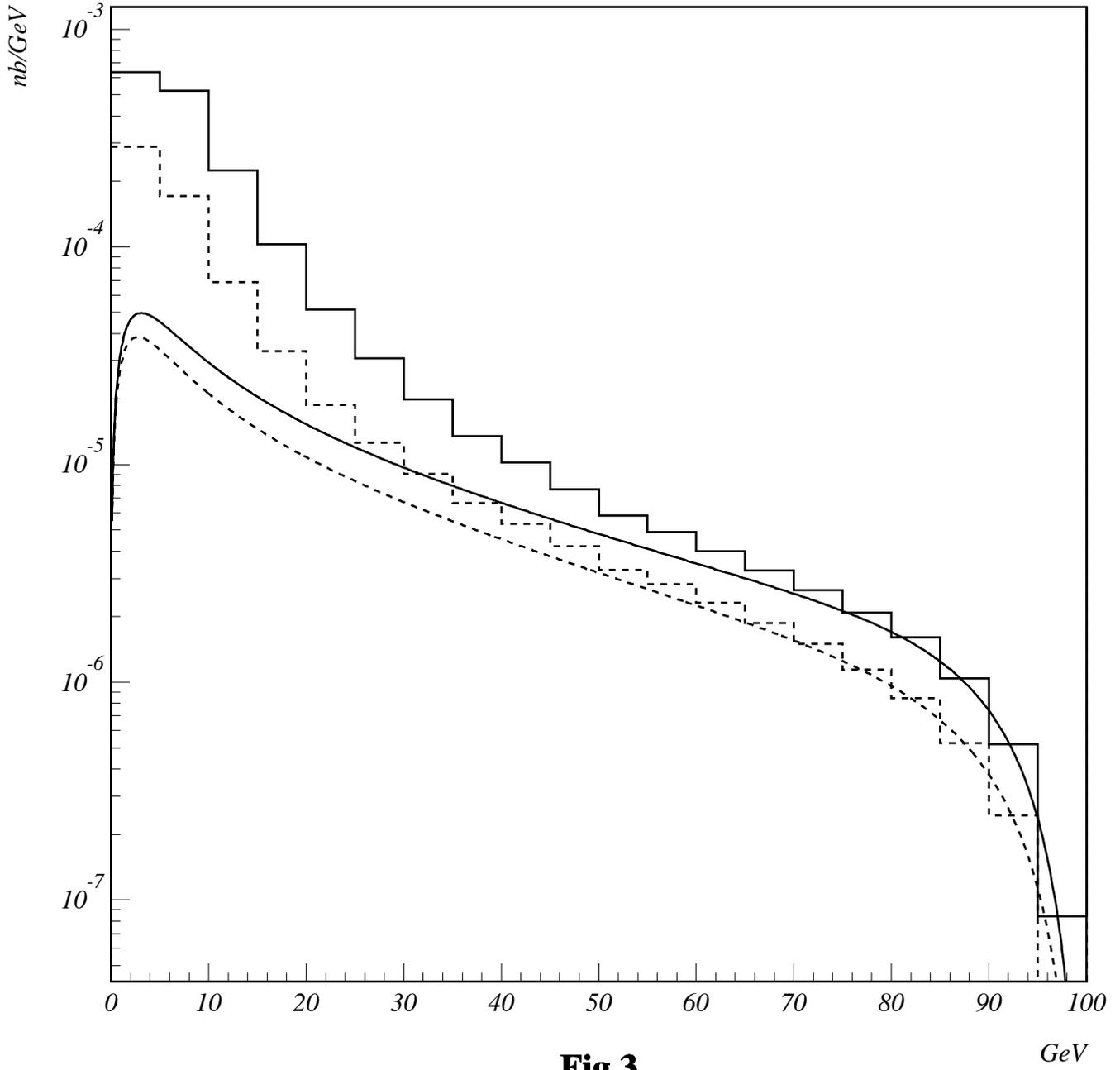
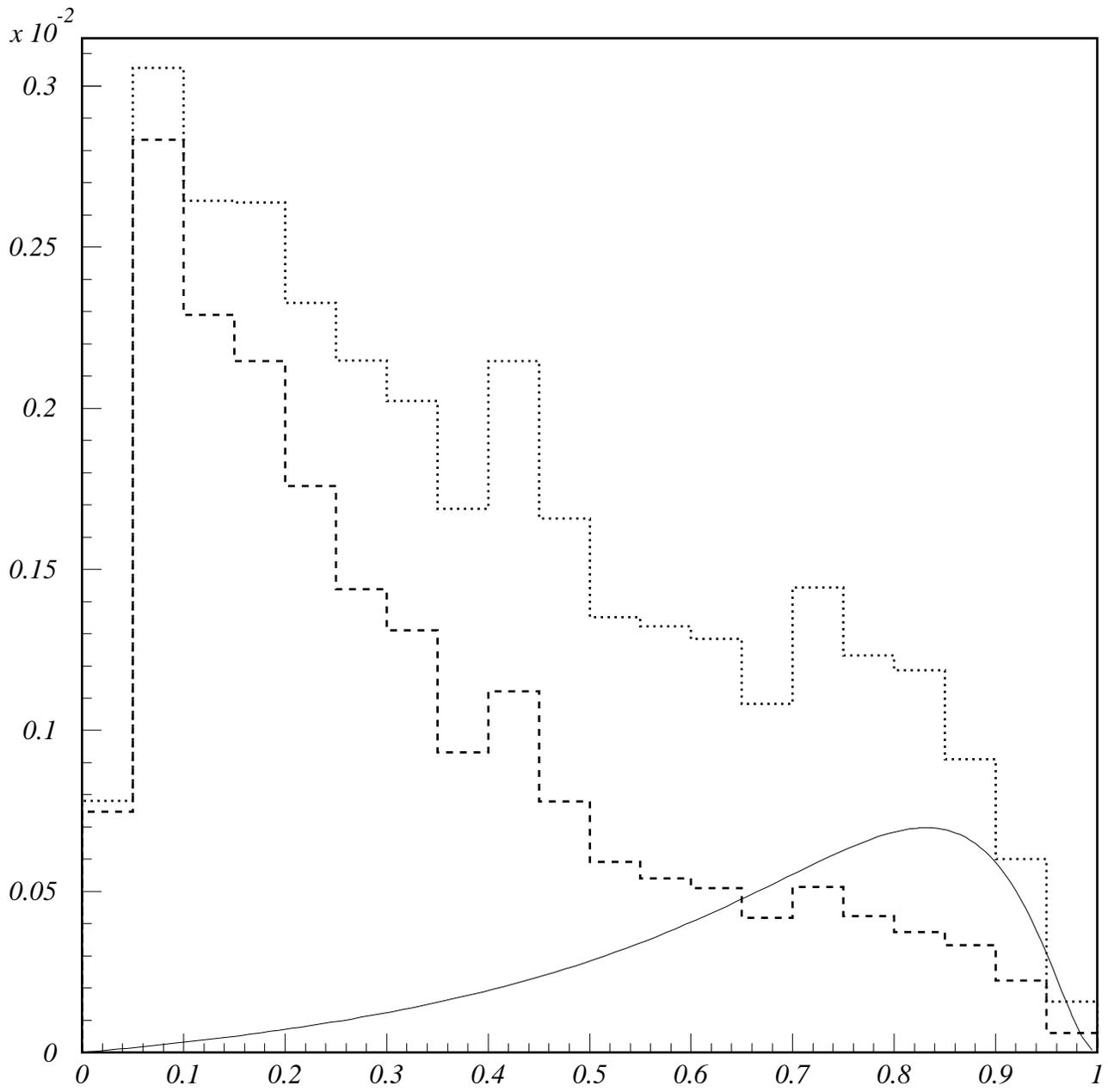
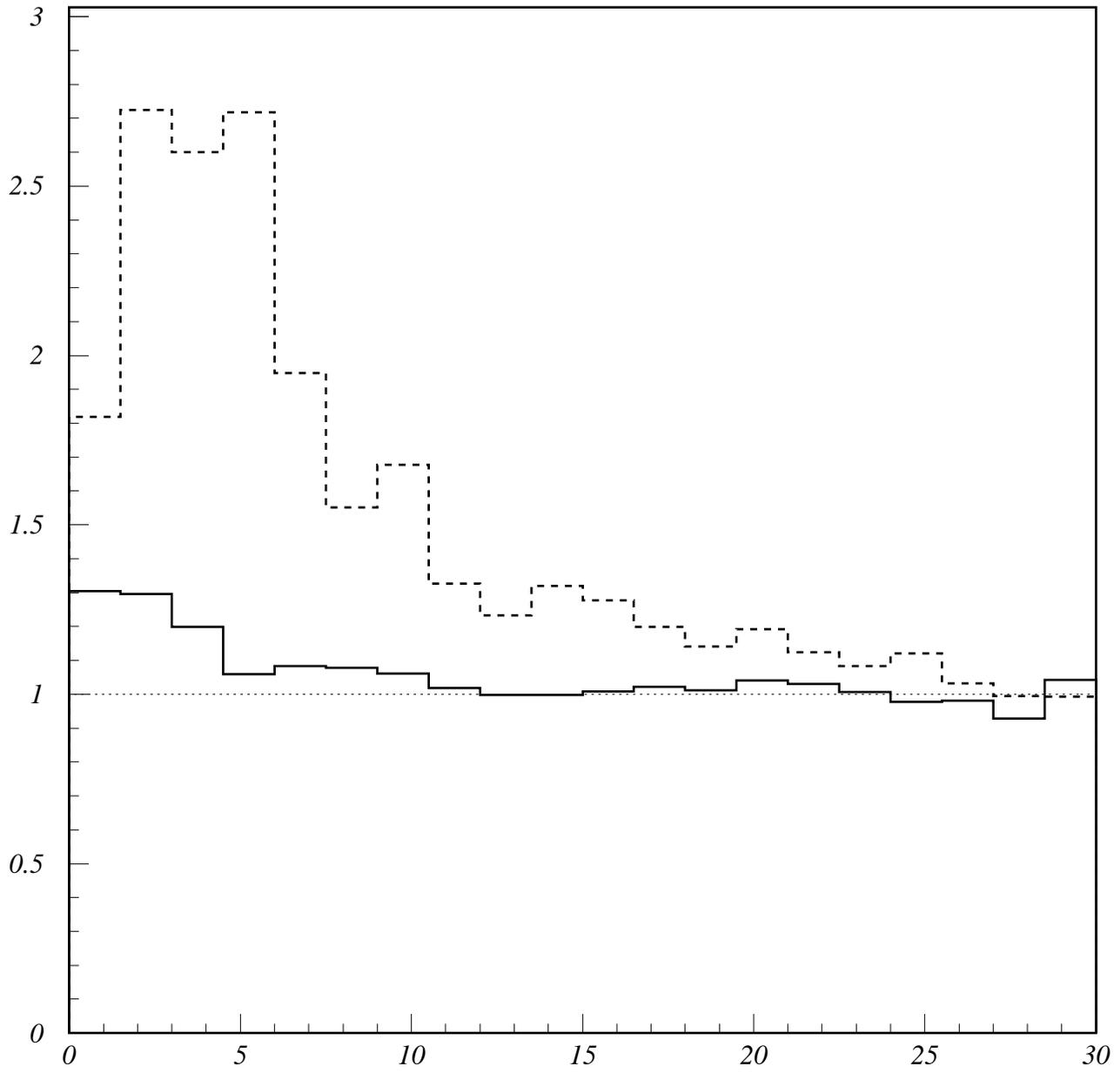
\end{document}